\documentclass[a4paper,prl,reprint,%
superscriptaddress,%
preprintnumbers,%
nobibnotes,%
amsmath,%
amssymb,%
aps]{revtex4-1}

\pdfoutput=1

\usepackage[utf8]{inputenc}

\newcommand{\GeV}{\ensuremath{\,\mathrm{GeV}}\xspace}

\usepackage{xspace}
\usepackage{graphicx}
\usepackage{color}
\usepackage{dcolumn}
\usepackage{bm}
\usepackage{color}

\begin{document}

\preprint{FTUV-18-0221, IFIC/18-06, KA-TP-04-2018, LU TP 18-03, UWTHPH-2018-4,
MCnet-18-04}

\title{Stress-testing the VBF approximation in multijet final states}

\author{Francisco Campanario}
\affiliation{Theory Division, IFIC, University of Valencia-CSIC, E-46980 Paterna, Valencia, Spain}
\affiliation{Institute for Theoretical Physics, Karlsruhe Institute of
  Technology, Germany}
\author{Terrance M. Figy}
\affiliation{Department of Mathematics, Statistics, and Physics, Wichita State
  University, Wichita, Kansas, USA}
\author{Simon Pl\"atzer}
\affiliation{Particle Physics, Faculty of Physics, University of Vienna, Austria}
\author{Michael Rauch}
\affiliation{Institute for Theoretical Physics, Karlsruhe Institute of
  Technology, Germany}
\author{Peter Schichtel}
\affiliation{German Research Center for Artificial Intelligence (DFKI),
  Kaiserslautern, Germany}
\affiliation{IAV Automotive Engineering, Kaiserslautern, Germany}
\author{Malin Sj\"odahl}
\affiliation{Department of Astronomy and Theoretical Physics, Lund University,
Sweden}

\begin{abstract}
  We consider electro-weak Higgs plus three jets production at NLO QCD beyond
  strict VBF acceptance cuts. We investigate, for the first time, how accurate
  the VBF approximation is in these regions and within perturbative
  uncertainties, by a detailed comparison of full and approximate
  calculations. We find that a rapidity gap between the tagging jets
  guarantees a good approximation, while an invariant mass cut alone is not
  sufficient, which needs to be confronted with experimental choices. We also
  find that a significant part of the QCD corrections can be attributed to
  Higgs-Strahlungs-type topologies.
\end{abstract}

\maketitle

\begin{figure}
\centering
\includegraphics[scale=0.55]{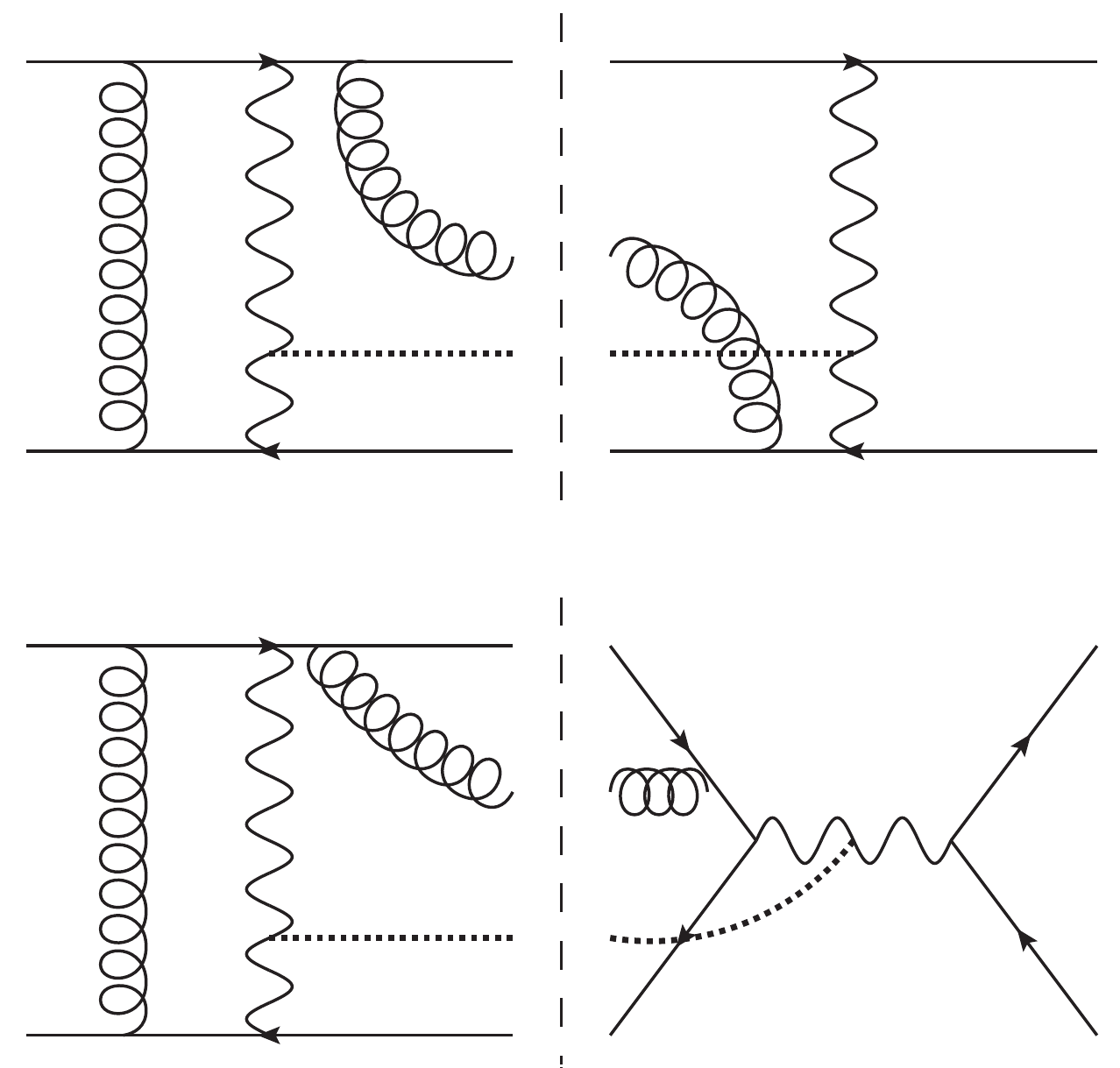}
\caption{\label{fig:diagrams}In the VBF region interferences among
  certain $t$-channel topologies (top diagram) as well as
  $t/s$-channel (lower diagram, and $t/u$-channel, not depicted)
  interferences are neglected. In these diagrams wavy lines represent
  electroweak bosons, dotted lines the Higgs boson, curly lines are
  gluons and solid arrowed lines are quarks. The dashed vertical line
  represents the final state cut.}
\end{figure}

{\it Introduction --} In 2012 both the ATLAS~\cite{Aad:2008zzm} and
CMS~\cite{Chatrchyan:2008aa} collaborations announced the discovery of
a new boson in the mass range of $125$--$126$
\GeV~\cite{Aad:2012tfa,Chatrchyan:2012xdj}. There were indications
that this new particle behaved very similar to the Higgs boson of the
Standard
Model~\cite{Kibble:1967sv,Higgs:1966ev,Guralnik:1964eu,Englert:1964et,Higgs:1964ia,Higgs:1964pj,Weinberg:1967tq,Glashow:1961tr}
and recently, the ATLAS and CMS collaborations reported the Standard
Model hypothesis to be consistent with data in a combined analysis of
LHC proton-proton collision data at $\sqrt{s}= 7$ and $8$
TeV~\cite{Khachatryan:2016vau}.  The Vector-Boson Fusion (VBF)
signature~\cite{Kauer:2000hi,Rainwater:1998kj,Rainwater:1997dg,Rainwater:1999sd,S.Asai:aa,Cranmer:2004uz,Eboli:2000ze,Ciccolini:2007jr,Ciccolini:2007ec,Bolzoni:2010xr,Bolzoni:2011cu,Cacciari:2015jma,Dreyer:2016oyx,Rauch:2016pai,Rauch:2017cfu,Cruz-Martinez:2018rod}
is among the most important production channels of the Higgs boson in
the ongoing run of the Large Hadron Collider (LHC).  For a Higgs boson
accompanied by at least two jets in the final state, the underlying
production processes allow for both space-like, $t$-channel, exchange
of weak gauge bosons producing a Higgs boson, as well as time-like
Higgs-{\it Strahlung} type topologies of associated production with a
vector boson which decays into a quark--anti-quark pair,
cf. Fig~\ref{fig:diagrams}. All of the contributing diagrams do
interfere, and the VBF region is usually referred to as a phase space
region in which one expects the $t$-channel diagrams to dominate, with
time-like $s$-channel and interference effects broadly suppressed.
In the VBF region one requires two highly energetic jets, well
separated in rapidity and with the Higgs boson decay products located
in the central detector region and possibly in between the two
jets. Additionally, a veto on central QCD activity is sometimes
applied to enrich the contribution of the colour singlet vector boson
exchange~\cite{Barger:1994zq,Rainwater:1996ud,Barger:1991ar,Forshaw:2007vb,Cox:2010ug,DuranDelgado:2011tp}.

Theoretical predictions in this region often employ the so-called VBF
approximation, where only the $t$-channel topology is kept and
$s$-channel contributions as well as interference effects between
different topologies are neglected. Fig~\ref{fig:diagrams} gives an
example of some of the contributions which are typically not
considered.  Formally, this corresponds to the approximation that the
constituents of the two incoming protons belong to two different, but
otherwise identical, copies of the colour gauge group $SU(3)$.  Recent
experimental analyses do not implement selection criteria for the VBF
region as tight as originally
envisaged~\cite{Kauer:2000hi,Rainwater:1998kj,Rainwater:1997dg,Rainwater:1999sd,S.Asai:aa,Cranmer:2004uz,Eboli:2000ze},
and rely on a multitude of multi-variate analysis techniques
instead~\cite{ATLAS:2016gld}. While for the Higgs plus two jet case
the validity of the VBF approximation has been confirmed within a
tight selection~\cite{Ciccolini:2007jr,Ciccolini:2007ec}, essentially
nothing is known quantitatively for additional radiation as relevant
to the veto on central jets (CJV), or virtually any observable
exploiting properties of the radiation pattern of the underlying
electro-weak production process.

Next-to-leading order corrections in Quantum Chromodynamics (QCD) to
the three jet process are available without any
approximation~\cite{Campanario:2013fsa} as a plugin to the
\textsf{Matchbox} framework~\cite{Platzer:2011bc} of the
\textsf{Herwig 7} event
generator~\cite{Bahr:2008pv,Bellm:2015jjp,Bellm:2017bvx}, and can be
compared to calculations based on the VBF
approximation~\cite{Figy:2007kv} as implemented in the \textsf{VBFNLO}
program~\cite{Arnold:2008rz,Arnold:2011wj,Baglio:2014uba}. In this
letter we quantify the reliability of the VBF approximation, {\it
  i.e.} the neglection of the diagrams which are not of the VBF
$t$-channel topology along with interference effects with $u$-channel
topologies.
\vspace*{1ex}

{\it Outline of the calculation --} We use the \textsf{Herwig 7} event
generator in its recent release 7.1.2~\cite{Bellm:2015jjp,Bellm:2017bvx},
together with \textsf{HJets++} 1.1 \cite{hjets} to provide the amplitudes for
electroweak Higgs boson plus jets production. The colour structure is treated
using ColorFull~\cite{Sjodahl:2014opa} and the loop integrals are computed
following Ref.~\cite{Campanario:2011cs}. For the VBF approximation we rely on
the approximate calculation provided by \textsf{VBFNLO} version 3.0 beta
5. Both calculations have recently also been interfaced to parton showers
using different matching paradigms, for a dedicated comparison
see~\cite{deFlorian:2016spz,Jager:2014vna}. The one-loop matrix elements of
\textsf{HJets++} and \textsf{VBFNLO} have been cross-checked against those of
\textsf{MadLoop}~\cite{Hirschi:2011pa}, \textsf{GoSam
  2.0}~\cite{Cullen:2014yla}, and \textsf{OpenLoops}~\cite{Cascioli:2011va} at
the level of phase space points.

We have ensured that both programs run with the same set of
electroweak parameters in a $G_\mu$ scheme with input parameters $G_F
= 1.16637 \times 10^{-5}\ {\rm GeV}^{-2}\ $, $M_Z = 91.1876\ {\rm
  GeV}$ and $ M_W = 80.403\ {\rm GeV}$.  The electromagnetic coupling
constant and the weak-mixing angle are calculated via tree level
relations. We take the Higgs boson as stable, with a mass fixed to
$m_H = 125.7\ {\rm GeV}$.  The widths of the bosons are fixed to
$\Gamma_Z=2.4952\ {\rm GeV}$ and $\Gamma_W=2.141\ {\rm GeV}$.  We
consider proton-proton collissions at $13\ {\rm TeV}$ center of mass
energy and employ a four-flavour scheme with the MMHT 2014 68\%
C.L. PDF set at NLO~\cite{Harland-Lang:2014aa} with a two-loop running
$\alpha_s$ set at $ \alpha_s(M_Z) = 0.12$ with $m_c = 1.4\ {\rm GeV} $
and $m_b = 4.75\ {\rm GeV}$.

We select jets using the anti-$k_\perp$ algorithm as implemented in the
\textsf{fastjet} library~\cite{Cacciari:2011aa,Cacciari:aa}, with a cone
radius of $R=0.4$, and accept jets ordered in transverse momentum, with a
transverse momentum $p_{\perp,j}> 30\ \GeV$ inside a rapidity range of
$|y_j|<4.4$. No restrictions are applied to the Higgs boson acceptance, nor
any other jet kinematic variable. We then use this baseline acceptance to scan
through possible cuts. Specifically, we consider tagging jet acceptances in
intervals of the leading dijet invariant mass $m_{12}=\sqrt{(p_{1}+p_{2})^2}$,
and the leading jet pair rapidity separation $\Delta y_{12}=|y_{1}-y_{2}|$,
\begin{eqnarray}
\label{eq:cuts}
m_{12} &>& m^{\text{cut}}_{12}\in \{0,100,200,300,400,500,600\}\ \GeV \, ,\nonumber\\
\Delta y_{12} &>& \Delta^{\text{cut}} y_{12}\in \{0,1,2,3\}\ .
\end{eqnarray}
The central renormalization, $\mu_R$, and factorization, $\mu_F$, scales are
chosen to be $H_\perp(\text{jets})$, which we here define as
\begin{equation}
  \label{eq:scale}
H_\perp =
  \frac{1}{2} \sum_{\substack{i\in \text{jets}\\\text{($p_{\perp} {>} 15\
  {\rm GeV}$)}}}
p_{\perp,i} \ ,
\end{equation}
where jets are clustered as outlined above, and only subject to a
reduced transverse momentum cut with $q_\perp=15\ {\rm GeV}$, which is
required to make the scale definition infrared and collinear
safe. Note that the jet cuts in the scale definition are more
inclusive than the analysis jet cuts.

The full calculation contains Higgs-{\it Strahlung} (VH) topologies,
which interfere with the possible VBF-type diagrams, as depicted in
Fig.~\ref{fig:diagrams}. While we expect these contributions not to be
relevant within tight VBF selection criteria, they might well
contribute when relaxing these constraints and as such yield a biased
view on quantifying the accuracy of the VBF approximation. Simulations
used by experimentalists also use a mix of VH and VBF processes, but
without interferences and without the pentagon
(Fig.~\ref{fig:diagrams}) and hexagon topologies, implying that biased
simulations go into experimental decisions and interpretation.

To work as closely as possible to the simulations used by
experimentalists we remove the VH contributions by applying a
resonance-veto on any single- and multi-jet masses in the neighborhood
of the $W^\pm$ and $Z$ masses, {\it i.e},
\begin{equation}
  \label{eq:resonance-veto}
  m_{V}-\delta m_{V} < m_{jets} < m_{V}+\delta m_{V}
\end{equation}
with $V=W^\pm,Z$. We choose $\delta m_{Z}=\delta m_{W}= 5~{\rm
  GeV}$. We discuss results both with, and without such a cut
applied. All analyses have been performed using a dedicated analysis
implemented in \textsf{Rivet}~\cite{Buckley:2010aa}.
\vspace*{1ex}

{\it Impact of QCD corrections --} For the inclusive selection, QCD
corrections have been found to be moderate for Higgs kinematics but
significant for third jet properties, specifically in the high-$p_{\perp,3}$
regime~\cite{Campanario:2013fsa}. The approximate calculation suggests small
corrections with a significant reduction in scale uncertainty. Prior to
studying the differences between the exact and the approximate calculations,
we have investigated the effect of QCD corrections subject to tight VBF cuts,
implemented by requesting a rapidity gap of $\Delta y_{12}>3$ and a invariant
mass of $m_{12}>600$ \GeV for two tagged jets.  We find that NLO corrections
in the VBF region are small, and the full and approximate calculations are in
reasonable agreement within $3$\%, with scale variations increasing by $8$\%
upon vetoing on resonant structures.

\begin{figure}[ht]
\centering
\includegraphics[scale=1.35]{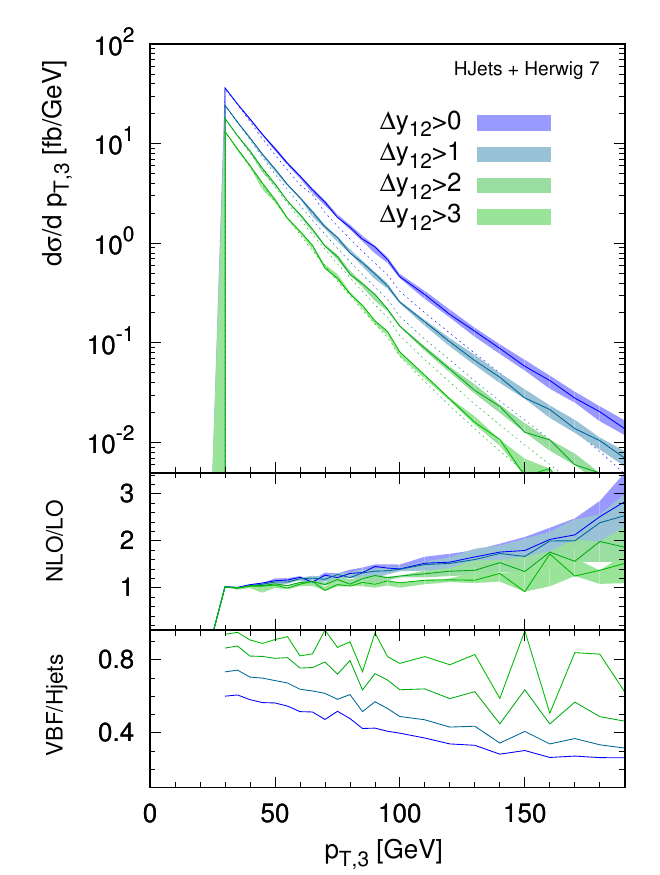}
\caption{\label{fig:jet3pt} QCD corrections on the third jet transverse momentum $p_{T,3}$
  spectrum for the full and approximate calculations for $m^{\rm
    cut}_{12}=0$\GeV and several choices of $\Delta^{\rm cut} y_{12}$
  (see eq.~\ref{eq:cuts}): the predictions for the full calculation,
  the corresponding differential $K$-factor and the ratio of the
  approximated over the full calculation are plotted in the top,
  middle and bottom panels, respectively.  We show LO results (dotted
  lines), NLO results (solid lines) with scale variations (light
  bands).}
\end{figure}

\begin{figure}
\centering
\includegraphics[scale=01.3]{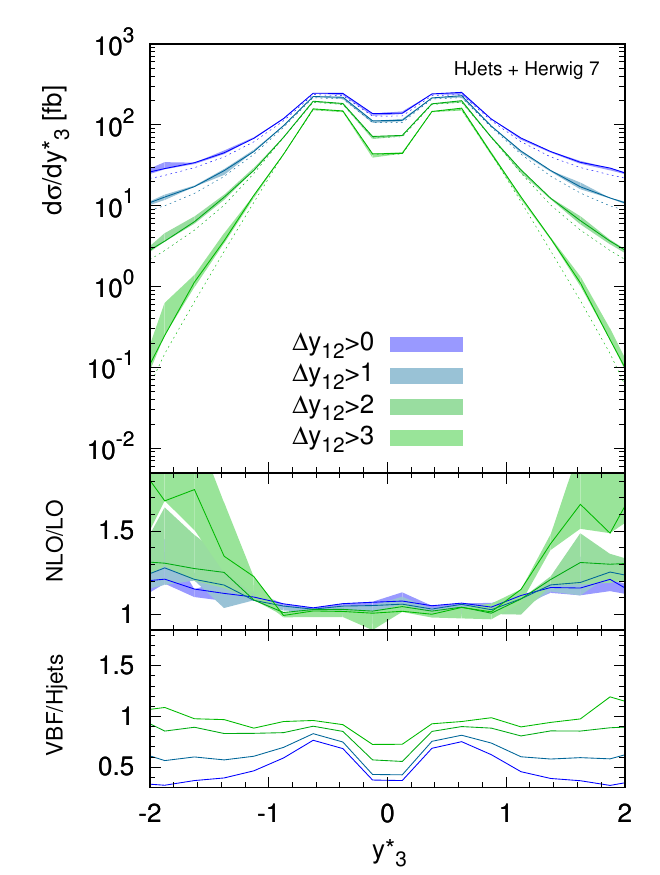}
\caption{\label{fig:y3} The normalized centralized rapidity $y_{3}^{\star}$
  distribution of the third jet for various leading jet separation
  rapidity cuts. The upper, middle and lower panel show respectively
  the full NLO and LO calculation (solid and dotted lines), the
  $K$-factor and the ratio of the VBF-approximation to the full NLO
  result.  }
\end{figure}

\begin{figure*}
\centering
\includegraphics[scale=1.3]{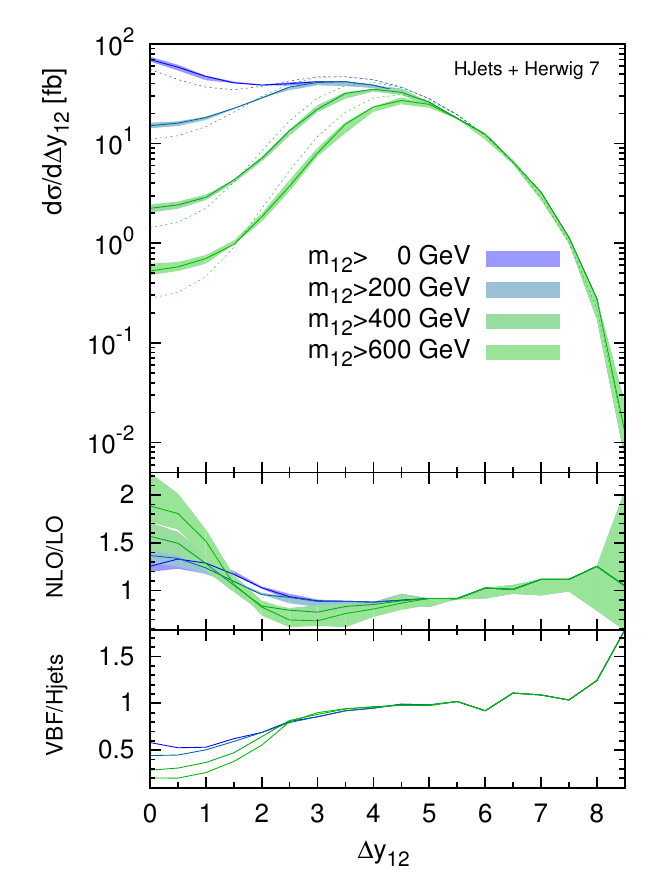}\hfill
\includegraphics[scale=1.3]{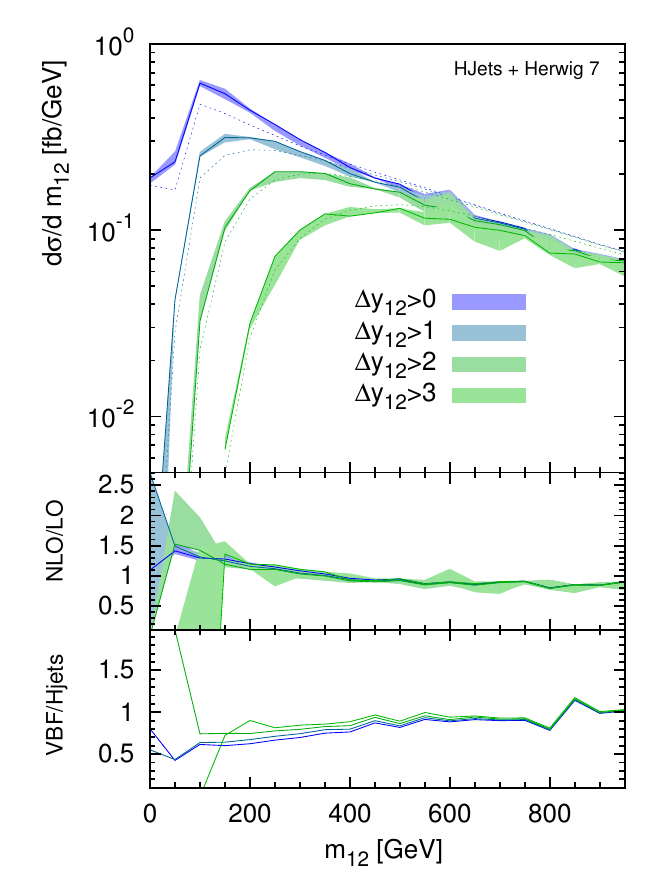}
\caption{\label{fig:dymjj} The rapidity separation $\Delta y_{12}$ of the leading two
  jets, for different cuts on their invariant mass (left) and the
  jet-jet invariant mass $m_{12}$ as a function of the rapidity gap requirement
  (right). We compare NLO QCD predictions in the full calculation
  (solid) to the approximate results (dashed). }
\end{figure*}

\begin{figure*}
\centering
\includegraphics[width=0.5\textwidth]{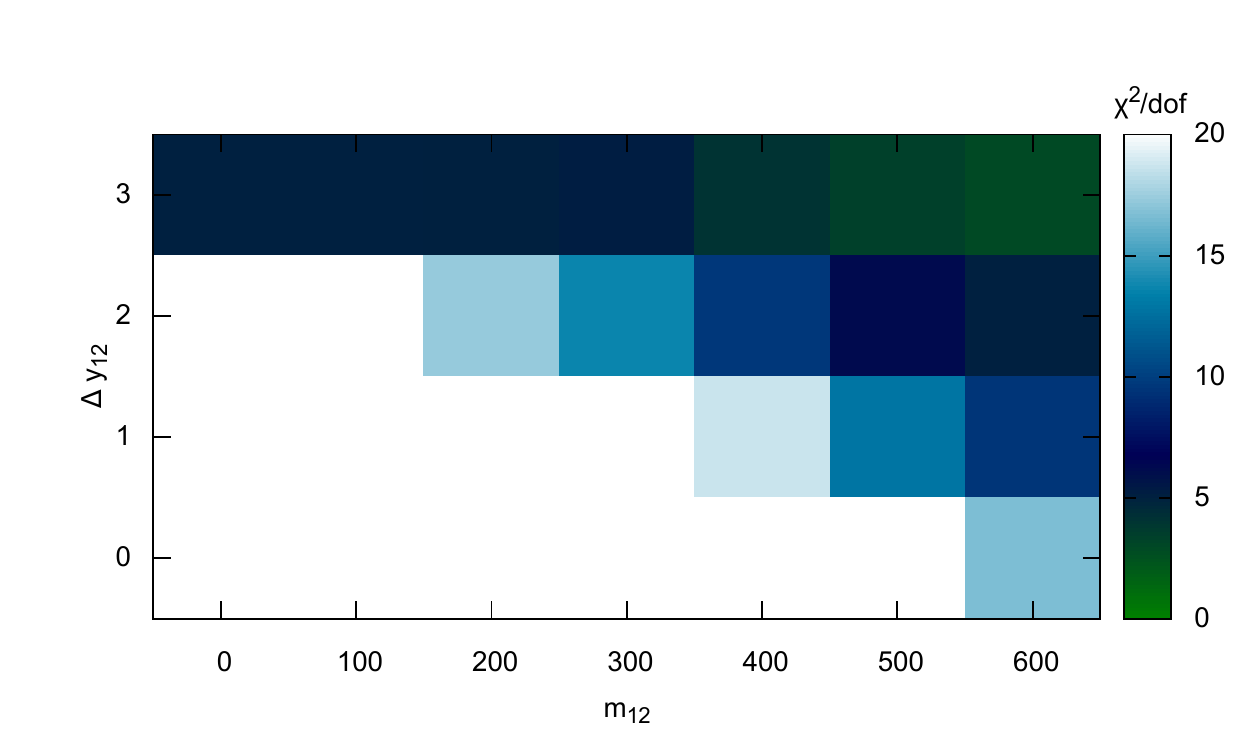}\hfill
\includegraphics[width=0.5\textwidth]{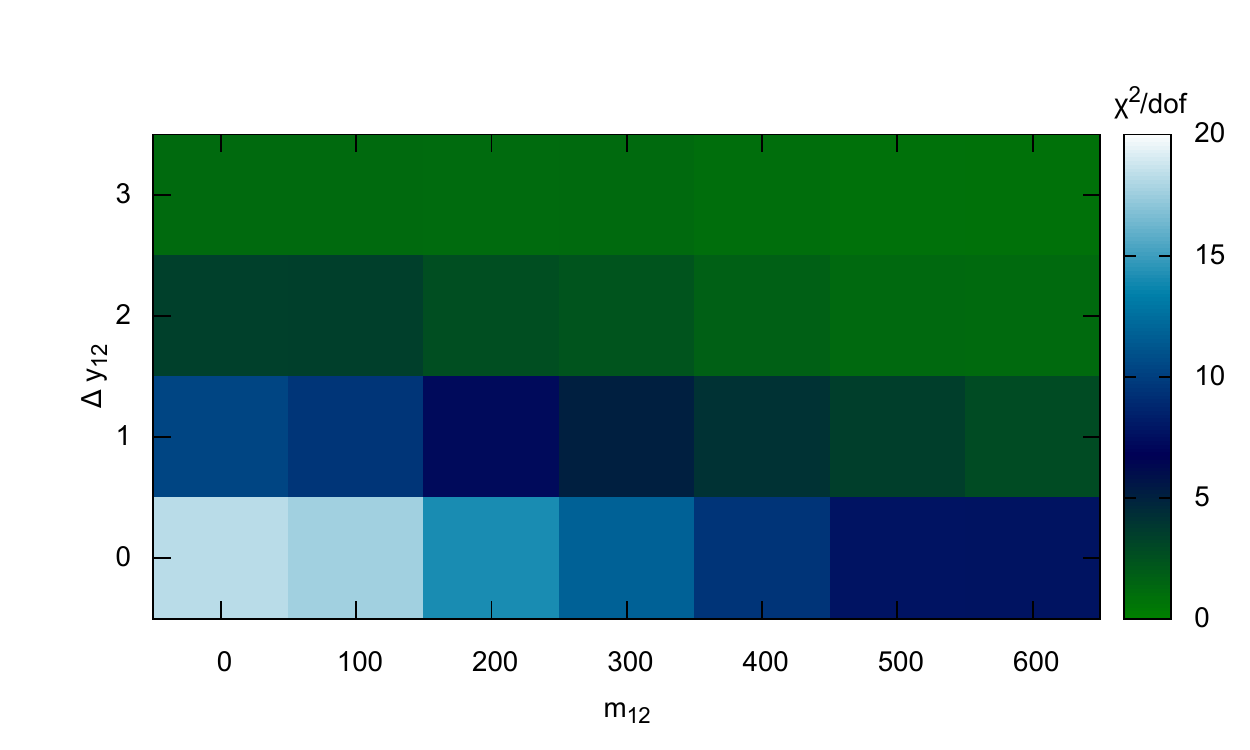}\\
\caption{\label{fig:chi2} Compatibility of the approximate predictions
  to the full calculation as a function of VBF acceptance cuts without
  (left) and with (right) applying vetos on Higgs-Strahlung-type
  contributions. We calculate goodness-of-fit measures based on the
  scale variation uncertainty for a range of observables relevant to
  the typical VBF kinematics. For the degrees of freedom we count only
  non-zero bins. Their number ranges from 101 to 129. }
\end{figure*}

Shown in the top panel of Fig.~\ref{fig:jet3pt} is the NLO (solid) and
LO (dotted) transverse momentum spectrum $p_{\perp,3}$ of the third
jet for the full calculation with $m^{\rm cut}_{12}=0$\GeV, no
resonance-veto cut applied and several choices of $\Delta^{\rm cut}
y_{12}$. The bands represent the NLO scale uncertainty in the range
$H_\perp/2 \le \mu_F=\mu_R \le 2 H_\perp$.

In all of the figures displaying differential cross sections, the
middle panel shows differential $K$-factors, defined as $ {\rm d}
\sigma_{\text{NLO}}/{\rm d}\sigma_{\text{LO}}$, where the bands
reflect the NLO scale variations with respect to the leading order
calculation fixed at the central scale. The increased $K$-factor
in the high transverse momentum region can be traced back to $VH+1$ jet
type events, and the resonance veto has the effect of reducing the
corrections down to values of $1.4$ in the high transverse momentum
region for the inclusive selection cuts ($\Delta y_{12}>0$) (not
shown). In the lower panel of Fig.~\ref{fig:jet3pt}, the ratio of the
approximate to the full result is plotted. Differences of order $50\%$
and more are visible when no rapidity separation is required.
However, as the rapidity gap increases, the large $K$-factor in the
transverse momentum spectrum for the full calculation is reduced
(cf.~middle panel), and the full and approximate results display
differences at the $20\%$ level in the bulk of the corrections for
$\Delta y_{12}>0$ (few to ten percent up to $p_{T,3}< 100$ GeV with a
resonance-cut applied), and increase up to $50\%$ ($30\%$ with a
resonance-cut) in the range shown.

In the upper panel in Fig.~\ref{fig:y3}, we consider the normalized
centralized rapidity distribution of the third jet $y_3^* =
(y_3-\frac{1}{2}(y_1+y_2))/|y_1-y_2|$ without resonance-veto cut for
the full NLO calculation as solid lines with scale uncertainty error
bands, as well as the LO result (dotted lines). QCD corrections tend
to increase for high rapidity separations $\Delta y_{12}$. We find a
clear improvement of the VBF-approximation for high rapidity
separations, whereas it will clearly underestimate the full result if
no rapidity separation is required.  This observation even holds when
resonance-cuts are applied (not shown) with differences in the central
and extreme regions of the plot of about $40\%$ for the $\Delta
y_{12}>0$ curve.

{\it Effects of $\Delta y_{12}$ and $m_{12}$ selections --}
Fig.~\ref{fig:dymjj} depicts the rapidity separation of the two
leading jets for several choices of $m^{\rm cut}_{12}$ (left) and the
dijet invariant mass of the leading two jets for several choices of
$\Delta^{\rm cut}y_{12}$ (right).  The resonance-veto described in
eq.~\eqref{eq:resonance-veto} has been enforced in the event selection
contributing to these observables.  We again compare NLO results
(solid lines) with scale uncertainty error bands for the full
calculation as well as LO results (dotted lines). NLO corrections
($K$-factors shown in the middle panels) can reach the $40\%$ level
for rapidity separation values $\leq 1$ and dijet invariant masses of
$100\ {\rm GeV}$.  Increasing the invariant mass cut $m_{12}$ beyond
approximately $100\ {\rm GeV}$ results in smaller NLO corrections for
all rapidity separations (as seen in the right plot of
Fig.~\ref{fig:dymjj}).  The quality of the VBF approximation is
shown in the lower ratio plots. Deviations of the order of several ten
per cent are visible for small rapidity separations and/or small dijet
invariant masses. Increasing values of $\Delta^{\rm cut} y_{12}$
result in better agreement between the full and approximate results
(left). However, the full and approximate calculations are not
guaranteed to agree in the presence of a cut on the dijet invariant
mass alone (right plot with $\Delta y >0$).

While we have so far only presented a few observables to quantify the
impact of QCD corrections and the validity of the VBF approximation,
the calculation we performed has actually involved a large number of
observables sensitive to the kinematic distribution of the third jet
as well as dedicated VBF observables. In order to quantify the quality
of the approximation across the whole set of these observables we
consider a metric inspired by a statistical test and calculate
\begin{equation}
\chi^2 = \frac{1}{N_{\text{bins}}}
\sum_{\text{bins }i}
  \frac{(\sigma_{i,\text{HJets}}-\sigma_{i,\text{VBF}})^2}{\text{
      max} \left(\delta_\mu\sigma_i^2,\delta_{\text{stat}}\sigma_i^2\right) \ ,
}
\end{equation}
where we consider the maximum of scale variation $\delta_\mu\sigma_i$ or
statistical deviation $\delta_{\text{stat}}\sigma_i$ per bin to set the scale
of fluctuations within which we want to measure agreement.  The results are
presented in Fig.~\ref{fig:chi2}, where we include $p_{T,3}, y_{3}^{\star},
y_{h}^{\star}, \Delta y_{h,12}, \Delta \phi_{h,12}$, and $m_{123}$ in the
goodness-of-fit calculation as a function of $\Delta^{\rm cut}y_{12}$ and
$m^{\rm cut}_{12}$ without (left column) and with (right column) the
resonance-veto on the Higgs-Strahlung-type events
Eq.~\eqref{eq:resonance-veto}. We can clearly observe that the VBF
approximation can be considered valid only for dijet invariant mass cuts above
$500$\GeV and for rapidity gaps above $2$. It would seem as if the VBF cuts do
not remove the $HVj$ events effectively even in tight VBF selections.  In
contrast, for the resonance-veto case agreement starts near $m_{12}=500$\GeV
and a rapidity gap of $0$, however only a rapidity gap cut of at least $2$
units guarantees decent agreement between the full and approximate
calculations.

\vspace*{1ex}

{\it Conclusions and outlook --} In this letter, we have addressed the
quality of the vector boson fusion approximation in three jet events by
comparing full and approximate calculations at NLO QCD. While moderate
rapidity separation cuts guarantee convergence at the percent level,
large dijet invariant mass cuts are not sufficient to achieve the
same accuracy. This important information should be taken into account
in experimental analyses. In addition, we have shown that the NLO QCD
corrections of the full calculation can reach a factor of 3 and are
consistent with Higgs-{\it Strahlung} $VHj$ contributions.

\vspace*{1ex}

{\it Acknowledgements --} FC acknowledges financial support by the Generalitat
Valenciana, Spanish Government and ERDF funds from the European Commission
(Grants No. RYC-2014-16061, SEJI-2017/2017/019,
FPA2017-84543-P,FPA2017-84445-P, and SEV-2014-0398). MR would like to
acknowledge the contribution of the COST Action CA16108. SP acknowledges
partial support by the COST Action CA16201 PARTICLEFACE. This work used the
Extreme Science and Engineering Discovery Environment (XSEDE), which is
supported by National Science Foundation grant number
ACI-1548562~\cite{xsede:6866038}. This work utilized computing resources
provided at IPPP, Durham. This work also utilized both the Open Science Grid
and TACC Stampede 2 to perform event generation runs through allocations
TG-TRA150015 and TG-PHY160001.  TF would like to thank Mats Rynge for
assistance with setting up event generation runs on the Open Science Grid,
which was made possible through the XSEDE Extended Collaborative Service
(ECSS) program. TF would like to thank Juan Cruz-Martinez, Jonas Lindert, and
Nigel Glover for valuable discussions. MR would like to thank Amon Engemann for
helpful code comparisons.

\bibliography{vbf-approximation}

\begin{thebibliography}{10}

\bibitem{Aad:2008zzm}
ATLAS, G.~Aad {\em et~al.},
\newblock JINST {\bf 3}, S08003 (2008).

\bibitem{Chatrchyan:2008aa}
CMS, S.~Chatrchyan {\em et~al.},
\newblock JINST {\bf 3}, S08004 (2008).

\bibitem{Aad:2012tfa}
ATLAS, G.~Aad {\em et~al.},
\newblock Phys. Lett. {\bf B716}, 1 (2012), 1207.7214.

\bibitem{Chatrchyan:2012xdj}
CMS, S.~Chatrchyan {\em et~al.},
\newblock Phys. Lett. {\bf B716}, 30 (2012), 1207.7235.

\bibitem{Kibble:1967sv}
T.~W.~B. Kibble,
\newblock Phys. Rev. {\bf 155}, 1554 (1967).

\bibitem{Higgs:1966ev}
P.~W. Higgs,
\newblock Phys. Rev. {\bf 145}, 1156 (1966).

\bibitem{Guralnik:1964eu}
G.~S. Guralnik, C.~R. Hagen, and T.~W.~B. Kibble,
\newblock Phys. Rev. Lett. {\bf 13}, 585 (1964).

\bibitem{Englert:1964et}
F.~Englert and R.~Brout,
\newblock Phys. Rev. Lett. {\bf 13}, 321 (1964).

\bibitem{Higgs:1964ia}
P.~W. Higgs,
\newblock Phys. Lett. {\bf 12}, 132 (1964).

\bibitem{Higgs:1964pj}
P.~W. Higgs,
\newblock Phys. Rev. Lett. {\bf 13}, 508 (1964).

\bibitem{Weinberg:1967tq}
S.~Weinberg,
\newblock Phys. Rev. Lett. {\bf 19}, 1264 (1967).

\bibitem{Glashow:1961tr}
S.~L. Glashow,
\newblock Nucl. Phys. {\bf 22}, 579 (1961).

\bibitem{Khachatryan:2016vau}
ATLAS, CMS, G.~Aad {\em et~al.},
\newblock JHEP {\bf 08}, 045 (2016), 1606.02266.

\bibitem{Kauer:2000hi}
N.~Kauer, T.~Plehn, D.~L. Rainwater, and D.~Zeppenfeld,
\newblock Phys. Lett. {\bf B503}, 113 (2001), hep-ph/0012351.

\bibitem{Rainwater:1998kj}
D.~L. Rainwater, D.~Zeppenfeld, and K.~Hagiwara,
\newblock Phys. Rev. {\bf D59}, 014037 (1998), hep-ph/9808468.

\bibitem{Rainwater:1997dg}
D.~L. Rainwater and D.~Zeppenfeld,
\newblock JHEP {\bf 12}, 005 (1997), hep-ph/9712271.

\bibitem{Rainwater:1999sd}
D.~L. Rainwater and D.~Zeppenfeld,
\newblock Phys. Rev. {\bf D60}, 113004 (1999), hep-ph/9906218,
\newblock [Erratum: Phys. Rev.D61,099901(2000)].

\bibitem{S.Asai:aa}
S.Asai {\em et~al.},
\newblock hep-ph/0402254.

\bibitem{Cranmer:2004uz}
K.~Cranmer, B.~Mellado, W.~Quayle, and S.~L. Wu,
\newblock (2004), hep-ph/0401088.

\bibitem{Eboli:2000ze}
O.~J.~P. Eboli and D.~Zeppenfeld,
\newblock Phys. Lett. {\bf B495}, 147 (2000), hep-ph/0009158.

\bibitem{Ciccolini:2007jr}
M.~Ciccolini, A.~Denner, and S.~Dittmaier,
\newblock Phys. Rev. Lett. {\bf 99}, 161803 (2007), 0707.0381.

\bibitem{Ciccolini:2007ec}
M.~Ciccolini, A.~Denner, and S.~Dittmaier,
\newblock Phys. Rev. {\bf D77}, 013002 (2008), 0710.4749.

\bibitem{Bolzoni:2010xr}
P.~Bolzoni, F.~Maltoni, S.-O. Moch, and M.~Zaro,
\newblock Phys. Rev. Lett. {\bf 105}, 011801 (2010), 1003.4451.

\bibitem{Bolzoni:2011cu}
P.~Bolzoni, F.~Maltoni, S.-O. Moch, and M.~Zaro,
\newblock Phys. Rev. {\bf D85}, 035002 (2012), 1109.3717.

\bibitem{Cacciari:2015jma}
M.~Cacciari, F.~A. Dreyer, A.~Karlberg, G.~P. Salam, and G.~Zanderighi,
\newblock Phys. Rev. Lett. {\bf 115}, 082002 (2015), 1506.02660.

\bibitem{Dreyer:2016oyx}
F.~A. Dreyer and A.~Karlberg,
\newblock Phys. Rev. Lett. {\bf 117}, 072001 (2016), 1606.00840.

\bibitem{Rauch:2016pai}
M.~Rauch,
\newblock (2016), 1610.08420.

\bibitem{Rauch:2017cfu}
M.~Rauch and D.~Zeppenfeld,
\newblock Phys. Rev. {\bf D95}, 114015 (2017), 1703.05676.

\bibitem{Cruz-Martinez:2018rod}
J.~Cruz-Martinez, T.~Gehrmann, E.~W.~N. Glover, and A.~Huss,
\newblock (2018), 1802.02445.

\bibitem{Barger:1994zq}
V.~D. Barger, R.~J.~N. Phillips, and D.~Zeppenfeld,
\newblock Phys. Lett. {\bf B346}, 106 (1995), hep-ph/9412276.

\bibitem{Rainwater:1996ud}
D.~L. Rainwater, R.~Szalapski, and D.~Zeppenfeld,
\newblock Phys. Rev. {\bf D54}, 6680 (1996), hep-ph/9605444.

\bibitem{Barger:1991ar}
V.~D. Barger, K.-m. Cheung, T.~Han, and D.~Zeppenfeld,
\newblock Phys. Rev. {\bf D44}, 2701 (1991),
\newblock [Erratum: Phys. Rev.D48,5444(1993)].

\bibitem{Forshaw:2007vb}
J.~R. Forshaw and M.~Sjodahl,
\newblock JHEP {\bf 09}, 119 (2007), 0705.1504.

\bibitem{Cox:2010ug}
B.~E. Cox, J.~R. Forshaw, and A.~D. Pilkington,
\newblock Phys. Lett. {\bf B696}, 87 (2011), 1006.0986.

\bibitem{DuranDelgado:2011tp}
R.~M. Duran~Delgado, J.~R. Forshaw, S.~Marzani, and M.~H. Seymour,
\newblock JHEP {\bf 08}, 157 (2011), 1107.2084.

\bibitem{ATLAS:2016gld}
ATLAS, T.~A. collaboration,
\newblock (2016).

\bibitem{Campanario:2013fsa}
F.~Campanario, T.~M. Figy, S.~Pl{\"a}tzer, and M.~Sj{\"o}dahl,
\newblock Phys. Rev. Lett. {\bf 111}, 211802 (2013), 1308.2932.

\bibitem{Platzer:2011bc}
S.~Platzer and S.~Gieseke,
\newblock Eur. Phys. J. {\bf C72}, 2187 (2012), 1109.6256.

\bibitem{Bahr:2008pv}
M.~Bahr {\em et~al.},
\newblock Eur. Phys. J. {\bf C58}, 639 (2008), 0803.0883.

\bibitem{Bellm:2015jjp}
J.~Bellm {\em et~al.},
\newblock Eur. Phys. J. {\bf C76}, 196 (2016), 1512.01178.

\bibitem{Bellm:2017bvx}
J.~Bellm {\em et~al.},
\newblock (2017), 1705.06919.

\bibitem{Figy:2007kv}
T.~Figy, V.~Hankele, and D.~Zeppenfeld,
\newblock JHEP {\bf 02}, 076 (2008), 0710.5621.

\bibitem{Arnold:2008rz}
K.~Arnold {\em et~al.},
\newblock Comput. Phys. Commun. {\bf 180}, 1661 (2009), 0811.4559.

\bibitem{Arnold:2011wj}
K.~Arnold {\em et~al.},
\newblock (2011), 1107.4038.

\bibitem{Baglio:2014uba}
J.~Baglio {\em et~al.},
\newblock (2014), 1404.3940.

\bibitem{hjets}
\texttt{http://hjets.hepforge.org}.

\bibitem{Sjodahl:2014opa}
M.~Sjodahl,
\newblock Eur. Phys. J. {\bf C75}, 236 (2015), 1412.3967.

\bibitem{Campanario:2011cs}
F.~Campanario,
\newblock JHEP {\bf 10}, 070 (2011), 1105.0920.

\bibitem{deFlorian:2016spz}
LHC Higgs Cross Section Working Group, D.~de~Florian {\em et~al.},
\newblock (2016), 1610.07922.

\bibitem{Jager:2014vna}
B.~J{\"a}ger, F.~Schissler, and D.~Zeppenfeld,
\newblock JHEP {\bf 07}, 125 (2014), 1405.6950.

\bibitem{Hirschi:2011pa}
V.~Hirschi {\em et~al.},
\newblock JHEP {\bf 05}, 044 (2011), 1103.0621.

\bibitem{Cullen:2014yla}
G.~Cullen {\em et~al.},
\newblock Eur. Phys. J. {\bf C74}, 3001 (2014), 1404.7096.

\bibitem{Cascioli:2011va}
F.~Cascioli, P.~Maierhofer, and S.~Pozzorini,
\newblock Phys. Rev. Lett. {\bf 108}, 111601 (2012), 1111.5206.

\bibitem{Harland-Lang:2014aa}
L.~A. Harland-Lang, A.~D. Martin, P.~Motylinski, and R.~Thorne,
\newblock (2014), 1412.3989.

\bibitem{Cacciari:2011aa}
M.~Cacciari, G.~P. Salam, and G.~Soyez,
\newblock (2011), 1111.6097.

\bibitem{Cacciari:aa}
M.~Cacciari and G.~P. Salam,
\newblock hep-ph/0512210.

\bibitem{Buckley:2010aa}
A.~Buckley {\em et~al.},
\newblock (2010), 1003.0694.

\bibitem{xsede:6866038}
J.~Towns {\em et~al.},
\newblock {Computing in Science and Engineering} {\bf 16}, 62 (2014).

\end{thebibliography}

\end{document}